\documentclass[useAMS,usenatbib,usegraphicx,a4paper]{mn2e}
\usepackage{times}
\newcommand{\aj}{AJ}                   
\newcommand{\apj}{ApJ}                 

\newcommand{\apss}{Ap\&SS}             
\newcommand{\aap}{A\&A}                
  
\newcommand{\aaps}{A\&AS}              
\newcommand{\mnras}{MNRAS}             
\newcommand{\pasp}{PASP}               

\newcommand{\echa}{$\eta$~Cha} 

\newcommand{\kms}{km~s$^{-1}$}
\newcommand{\msun}{M$_{\sun}$}
\newcommand{\twentynine}{2MASS 0801--8058}
\newcommand{\thirtytwo}{2MASS 0820--8003}
\newcommand{\fortyone}{2MASS 0913--7550}
\newcommand{\fiftyone}{2MASS 0905--8134}
\newcommand{\sixtyfour}{2MASS 0955--7622}
\newcommand{\fiftyeight}{2MASS 0942--7727}

\begin{document}
\title[First detection of a low-mass stellar halo around $\eta$~Cha]{First detection of a low-mass stellar halo around the young open cluster $\eta$~Chamaeleontis}
\author[S. J. Murphy, W. A. Lawson and M. S. Bessell]{Simon~J.~Murphy$^1$\thanks{Email: murphysj@mso.anu.edu.au (SJM); w.lawson@adfa.edu.au (WAL); bessell@mso.anu.edu.au (MSB)}, Warrick~A.~Lawson$^2$\footnotemark[1] and Michael~ S.~Bessell$^1$\footnotemark[1] \\
$^1$ Research School of Astronomy and Astrophysics, The Australian National University, Cotter Road, 
Weston Creek ACT 2611, Australia \\
$^2$  School of PEMS, University of New South Wales, Australian Defence Force Academy, Canberra, ACT 2600, Australia }
\maketitle
\begin{abstract}
We have identified several lithium-rich low-mass ($0.08<M<0.3$ \msun) stars within 5.5 deg of the young open cluster $\eta$ Chamaeleontis, nearly four times the radius of previous search efforts. Of these stars we propose 4 new probable cluster members, and 3 possible members requiring further investigation. These findings are consistent with a dynamical origin for the current configuration of the cluster, without the need to invoke an abnormal Initial Mass Function deficient in low-mass objects. Candidates were selected on the basis of DENIS and 2MASS photometry, NOMAD astrometry and extensive follow-up spectroscopy. 
 \end{abstract}

\section{Introduction}
\label{sec:intro}

The recently-discovered open cluster $\eta$ Chamaeleontis is one of the closest ($d\sim94$~pc) and youngest ($t\sim8$~Myr) stellar associations in the Solar neighbourhood. A census of its stellar population currently stands at 18 systems, covering spectral types B8--M5.5. At high and intermediate masses the cluster Initial Mass Function (IMF) follows that of other star-forming regions and young stellar groups, but there is a clear deficit of members at masses $<$0.15 \msun. Comparing the observed mass function of the cluster to the Orion Trapezeium Cluster, \citet{Lyo04} predict that an additional 20 stars and brown dwarfs in the mass range $0.025<M<0.15$ \msun\ remain to be discovered. Efforts to observe this hitherto unseen population have failed to find any additional members at either larger radii from the cluster core \citep[][to 1.5~deg, 4 times the radius of known membership]{Luhman04} or to low masses in the cluster core \citep[][to $\sim$13~M$_{\rm Jup}$]{Lyo06}. Failure to find these low--mass members raises a fundamental question: has the cluster's evolution been driven by dynamical interactions which dispersed the stars into a diffuse halo at even larger radii, or does \echa\ possess an abnormally top-heavy IMF deficient in low--mass objects? The latter result would seemingly be at odds with the growing body of evidence that suggests the IMF is universal and independent of initial star-forming conditions \citep*[for a detailed review see][]{Bastian10}.

\citet*{Moraux07} have attempted to model the observed properties of \echa\ using $N$-body simulations of the cluster's dynamical evolution starting with standard initial conditions. They are able to replicate the current configuration of the cluster assuming a log-normal IMF and 30--70 initial members. This suggests the deficit of low--mass objects seen in the present day cluster may not be due to a peculiar IMF but to dynamical evolution. Their simulations predict there should exist a diffuse halo of cluster ejectees beyond the radius currently surveyed. Such comoving objects will have similar proper motions and distances as the cluster proper. In this letter we present the results of our search for this putative halo of low--mass stars surrounding \echa. 

\section{Candidate Selection}
\label{sec:candselect}

\subsection{Photometry}
\label{sec:photometry}

\begin{figure}
\includegraphics[width=0.45\textwidth]{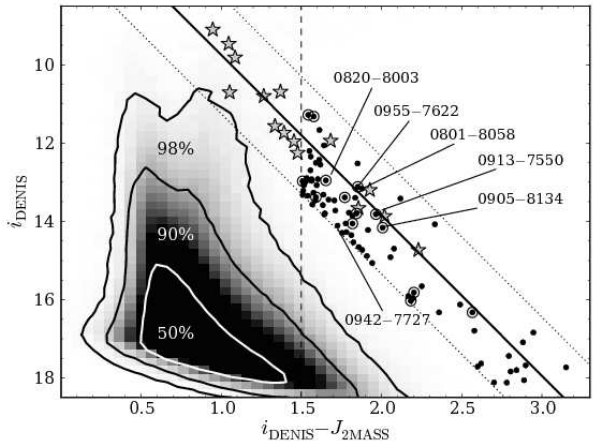} 
\caption{DENIS/2MASS colour--magnitude diagram for a 5.5 deg radius around \echa. Contours show the cumulative totals of stars enclosed. We select candidates (filled circles) $\pm$1.5~mag from the empirical cluster isochrone (solid line) also having $(i-J)>1.5$. Known KM-type members are shown as filled stars. Proper motion candidates are denoted by open circles. The intermediate--gravity stars described in \S\ref{sec:gravity} are labelled.}
\label{fig:cmd}
\end{figure}

\begin{figure} 
   \centering
   \includegraphics[width=0.45\textwidth]{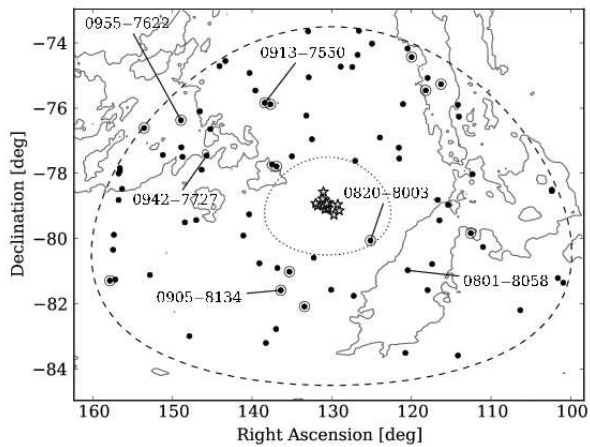} 
   \caption{Distribution of candidates on the sky. Symbols are as for Figure~\ref{fig:cmd}. The area enclosed by the dotted line is that  searched by \citet{Luhman04}, the dashed line is the 5.5~deg radius of this study. Contours correspond to 10~MJy~sr$^{-1}$ IRAS 100 $\micron$ flux. 
 }
   \label{fig:skyplot}
\end{figure}

Using the \textsc{Topcat}\footnote{Available for download at http://www.starlink.ac.uk/topcat/}
Virtual Observatory tool we crossmatched DENIS and 2MASS survey photometry within a 5.5 deg radius surrounding \echa. 
Candidates were selected by forming the $i_{\rm{DENIS}}$ versus $(i- J_{\rm{2MASS}})$ colour-magniude diagram for the $1.2\times10^6$ sources in the field. As seen in Figure~\ref{fig:cmd} the empirical cluster isochrone is well defined in this diagram through the K and M spectral types. Due to its youth, proximity to Earth and lack of any substantial reddening, the cluster isochrone sits well above the vast majority of stars in the field. To account for unresolved binarity, reddening, distance variations and any non-linearities in the isochrone we initially selected photometric candidates lying $\pm$1.5~mag from a linear fit to all known late-type members. Because we are interested in the low--mass population we restricted our candidates to $(i-J)>1.5$, corresponding to spectral types $\sim$M3 and later, and $(J-H)_{\rm{2MASS}}<0.8$, to minimize contamination from background giants. 
Excluding stars with poor 2MASS photometry (14), known members (5) and previously confirmed field stars (3) left 81 photometric candidates. 
Figure~\ref{fig:skyplot} shows the distribution of the candidates on the sky. Just one candidate lies within 1.5 deg of the cluster, the radius to which \cite{Luhman04} searched using a different method and reported no new members. 
 
\subsection{Proper Motions}
\label{sec:pm}

Kinematics have been vital in confirming membership of the numerous young stellar associations in the Solar neighborhood. In the absence of radial velocities and accurate distances however, proper motions are the sole means of investigating the kinematics of our stars. We therefore cross matched the 81 candidates against the Naval Observatory Merged Astrometric Catalogue (NOMAD). 
Only one of the photometric candidates was not found in NOMAD, this is the second faintest star in the sample. Eighteen candidates returned a null proper motion. These correspond to highly uncertain proper motion measurements explicitly set to zero (D.\ Monet, private communication). All 18 \echa\ members were found in NOMAD, 14 of which have non--zero proper motions. The four outstanding members are located close to bright stars or diffraction spikes, making proper motion measurements difficult. 

For a given $UVW$ space motion the resultant proper motion vector (and radial velocity) depends on both sky position and distance. Given the large angular extent of the survey area, this can have a substantial effect on the expected proper motions and radial velocities of our candidates. Using the canonical cluster space motion, $(U,V,W) = (-11.8,-19.1,-10.5)$ \kms\ \citep{Mamajek00} we calculated the expected proper motion of each candidate. From the uncertainties in the NOMAD proper motions we then calculated the number of sigma from the expected proper motion in each component. We selected as proper motion candidates the 14 stars that lay within $\pm3\sigma$ of the mean of the cluster stars. All 14 known members fell within the 3$\sigma$ selection box.

\section{Follow--up spectroscopy}
\label{sec:spectroscopy}

\subsection{Low--resolution spectroscopy}
\label{sec:lowres}

We obtained low--resolution spectra of our proper motion candidates and a selection of photometric candidates using the Double Beam Spectrograph (\textit{DBS}) on the ANU 2.3 m telescope at Siding Spring Observatory during 2009 March. With the 316~lines~mm$^{-1}$ red grating our spectra cover the wavelength range 5920--9750~\AA\ at a resolution of 5 \AA\ (1.9 \AA\ pixel$^{-1}$). On each night we also observed flux standards, smooth spectrum stars for telluric correction and a selection of Gliese dwarf standards from \cite{Lyo04a, Lyo08} for spectral typing and gravity estimation. Of the 14 proper motion candidates, 11 were observed in this mode. The remaining three are faint and two have DENIS $i$ magnitudes that fall more than 1 mag below the cluster isochrone and are not considered likely members. We also observed 18 additional photometric candidates that lay close to the cluster isochrone and had null proper motions, or a proper motion just outside the 3$\sigma$ selection box.

\subsection{Spectral types}
\label{sec:sptypes}

To garner spectral types we first performed synthetic photometry on the flux-calibrated \textit{DBS} spectra to calculate broadband $(R-I)$ colours. Unfortunately, the Gliese standards we intended to use to calibrate the colour transformations are much brighter than the candidates and suffer from severe fringing at wavelengths greater than $\sim$8000 \AA, probably as a result of scattered light inside the spectrograph. Using the Gliese dwarfs to calibrate the candidate photometry results in a systematic 0.05--0.15 mag over-estimation of $(R-I)$ for the known \echa\ members. We therefore calibrated the photometry of the Gliese dwarfs and candidates separately. The instrumental $(R-I)$ colours of the two groups were transformed onto the Cousins system using published photometry for the Gliese dwarfs  \citep{Bessell90} and the 14 late--type members of \echa\ from \citet{Lyo04a,Lyo08}, respectively. In both cases the transformation was well-fitted by a low-order polynomial. The \citeauthor{Lyo04a} \echa\ colours are calibrated using the same Gliese dwarfs and \citet{Bessell90} colours we used, so the two calibrations are equivalent.

Spectral types of the candidates were finally obtained by converting the $(R-I)$ colour to a spectral type using the M-dwarf transformation of \citet{Bessell91}. \citet{Lyo04a,Lyo08} have shown that the $(R-I)$ colours (and various narrowband spectral indices) of intermediate--age pre-main sequence (PMS) stars are indistinguishable from Gyr-old field dwarfs and that they closely follow a dwarf temperature (spectral type) sequence. The $(R-I)$ derived spectral types of the Gliese M-dwarfs and known \echa\ members agree with published values at the 0.1--0.2 subtype level. No attempt has been made to correct the photometry for the effects of reddening -- we assume it to be minimal at these distances. The maximum reddening from dust maps is $E(B-V)=0.5$~mag, although the majority of this material is probably background to the candidates. We estimate  0.1~mag as a more likely upper limit. 
To include the effects of unknown reddening we adopt 0.5 subtypes as the uncertainty on our spectral type measurements. 

\subsection{Gravity estimation and H$\alpha$ emission}
\label{sec:gravity}

\begin{figure}
   \centering
   \includegraphics[width=0.44\textwidth]{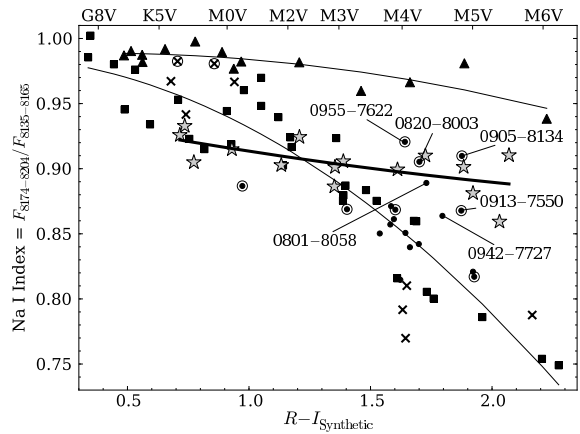} 
   \caption{Gravity--temperature diagram for our \textit{DBS} targets. Thin lines are fits to the observed K and M--type dwarfs (squares) and Pickles giants (triangles). The thick line is a fit to the late--type cluster members (filled stars). Candidates with H$\alpha$ emission are given by filled circles, those without are crosses. Proper motion candidates are denoted by open circles.}
   \label{fig:gravity_rmi}
\end{figure}

Pre--main sequence stars are in the process of contraction to their eventual main sequence radii. As a result they have slightly lower surface gravities than their main sequence counterparts (but still much closer to dwarfs than giants). The neutral alkali metal lines in our spectra, such as the K~I (7665/7699~\AA) and Na~I (8183/8195~\AA) doublets, are highly sensitive to gravity. Later than mid-M spectral types there is a marked difference in Na~I absorption between M--dwarfs, PMS stars and M--giants. Many authors have used the strength of the Na I doublet to gauge the surface gravities of late--type PMS candidates and assign coarse age rankings \citep[e.g.][]{Lyo04a,Lyo08,Lawson09b}. We adopt the spectral index defined by \citet{Slesnick06} to measure the strength of Na~I absorption. It is formed by integrating the spectrum in two 30 \AA\ bands, one on the doublet, the other sampling the adjacent pseudo-continuum. Figure~\ref{fig:gravity_rmi} shows the Na~I index versus $(R-I)$ colour for our \textit{DBS} targets, known \echa\ members, the Gliese field dwarfs and KM giants from the Pickles library. At spectral types later than M3 the locus of \echa\ members clearly diverges from the dwarf sequence and assumes an intermediate gravity between dwarfs and giants. Several candidates occupy the same region of the diagram -- these 6 stars are likely PMS stars of ages 5--10 Myr. 

All of the intermediate-gravity stars exhibit H$\alpha$ emission, as do all the late--type members of \echa. This is a necessary, but insufficient indicator of youth because of the potential contamination from older, flaring M-dwarfs. When combined with gravity information however it is a powerful discriminant. The six intermediate--gravity stars and their particulars are presented in Table~\ref{table:cand}

\begin{table*} 
\centering
\begin{minipage}{\textwidth} 
\caption{Candidate $\eta$ Chamaeleontis members in the surveyed region} 
\label{table:cand}
\begin{tabular}{lccccccccc} 
\hline
2MASS & $i_{\rm{DENIS}}$ & $(i-J)$ & Sp. Type. & H$\alpha$ EW & Li I EW & $V_{R}$ & $\mu_{\alpha}$ & $\mu_{\delta}$ & Membership? \\
identifier &  & & $\pm$0.5 & [$\pm$1 \AA] & [$\pm$ 0.05 m\AA] & [$\pm$ 2 km s$^{-1}]$& \multicolumn{2}{c}{[$\pm$ 9 mas yr$^{-1}$]} & (reason) \\
\hline
J08014860--8058052 & 13.17 & 1.88 & M4.4 & $-$20$<$EW$<$$-$7 & 0.55 & 19.3 & $-$28 & $+$39 & possible (binary?)  \\
J08202975--8003259 & 12.96 & 1.65 & M4.3 & $-$40$<$EW$<$$-$15 & 0.60 & 18.4 &$-$33 & $+$30& probable\\
J09053087--8134572 & 14.16 & 2.01 & M4.9 & $-$9  & 0.65  & 16.2& $-$35 & $+$27 & probable\\
J09133435--7550099 & 13.82 & 1.97 & M4.8 & $-$8 & 0.70 &12.9 &$-$40 &$+$34 & probable\\
J09424157--7727130 & 14.11 & 1.73 & M4.6 & $-$7 & 0.35 & 18.3 &$-$16 & $+$23 & no (Li, CMD)\\
J09553919--7622119 & 13.12 & 1.85 & M4.1 & $-$5 & 0.60 & 8.8 & $-$30&$+$19 & possible (dynamics)\\
\hline
\multicolumn{5}{l}{\textit{\citet{Covino97} ROSAT candidates:}}\\
RX J0902.9$-$7759 & 11.55 & 1.45 & M3 & $-$2  & 0.55  & 18.7& $-$35 & $+$31 & probable\\
RX J0915.5$-$7608 & 10.44 & 1.11 & K7 & $-$1 & 0.55 &22.0 &$-$32 &$+$16 &  no (too far)\\
RX J0942.7$-$7726 & 11.49 & 1.12 & M0 & $-$3 & 0.45 & 19.1 &$-$24 & $+$17 & no (Li, CMD) \\
RX J1005.3$-$7749 & 11.17 & 1.36 & M1 & $-$4 & 0.60 & 17.7 & $-$30&$+$17 & possible (dynamics)\\
\hline
\end{tabular} 
\end{minipage} 
\end{table*} 

\subsection{Medium--resolution spectroscopy}
\label{sec:medres}

To derive radial velocities and check for the presence of lithium absorption at 6708 \AA, we obtained medium--resolution ($R\simeq7000$) spectra of the stars in Table~\ref{table:cand} with the \textit{WiFeS} instrument at the ANU 2.3~m during 2010 January--April. We also observed many of the other candidates with congruent photometry, proper motions or H$\alpha$ emission. \textit{WiFeS} \citep{Dopita07} is a double--beam, image--slicing integral field spectrograph, and provides a 25$\times$38~arcsec field with 0.5$''$ pixels. The red $R$7000 grating provides wavelength coverage from 5300--7060~\AA, at a resolution of 1~\AA\ (0.44~\AA~px$^{-1}$). 
For deriving radial velocities, we took NeAr arc frames immediately after each exposure and also observed 5--7 mid-M radial velocity standards each night. Only the 6 intermediate--gravity stars in Table~\ref{table:cand} were found to have lithium in their spectra. To obtain a radial velocity each spectrum was cross-correlated against all the standards from that night over the region 6000--6500~\AA, and the velocity taken as the mean of those measurements (after correcting to a heliocentric frame). Stars were observed on multiple nights and we took the velocity as the mean of these values. Multiple measurements and cross-correlations between standards show we are easily able to attain a radial velocity precision of 2~km~s$^{-1}$ from the instrument.  Li~I and H$\alpha$ equivalent widths and the radial velocities derived from the \textit{WiFeS} spectra are given in Table~\ref{table:cand}.

\section{Cluster membership}
\subsection{Lithium equivalent widths}
\label{sec:lithium}

Lithium is very easily destroyed in stellar atmospheres, hence the detection of lithium absorption in a low--mass star is a strong indicator of youth. The equivalent width of the Li~I 6708 \AA\ feature is highly dependent on both age and temperature. \citet{Mentuch08} present the distribution of Li~I EWs for a variety of nearby stellar associations. For the spectral types under consideration here all but one of the measured Li~I EWs in Table~\ref{table:cand} are consistent with an 8--10~Myr population like \echa.  Based on its smaller Li~I equivalent width and poor CMD placement we are able to immediately rule out \fiftyeight\ as an \echa\ member. 

\subsection{Other young stars in the region}

The region surrounding \echa\ is rich in pre-main sequence stars of various ages. \citet{Covino97} present radial velocities, spectral types and lithium equivalent widths from high-resolution spectra of ROSAT detections in the region. Four of their stars lie in the eastern half of our survey area and have spectral types and lithium equivalent widths similar to the late-type population of \echa. These are shown in Table~\ref{table:cand} with measurements derived from \textit{WiFeS} spectra. Spectral types are from \citet{Covino97}. Such an X-ray selected sample is probably incomplete, as the ROSAT flux limit for a $\sim$10 Myr PMS star at the cluster distance is around M2. 

\subsection{Dynamics}
\label{sec:dynamics}

\begin{figure*}
   \centering
   \begin{tabular}{ccc}
   \includegraphics[width=0.31\textwidth]{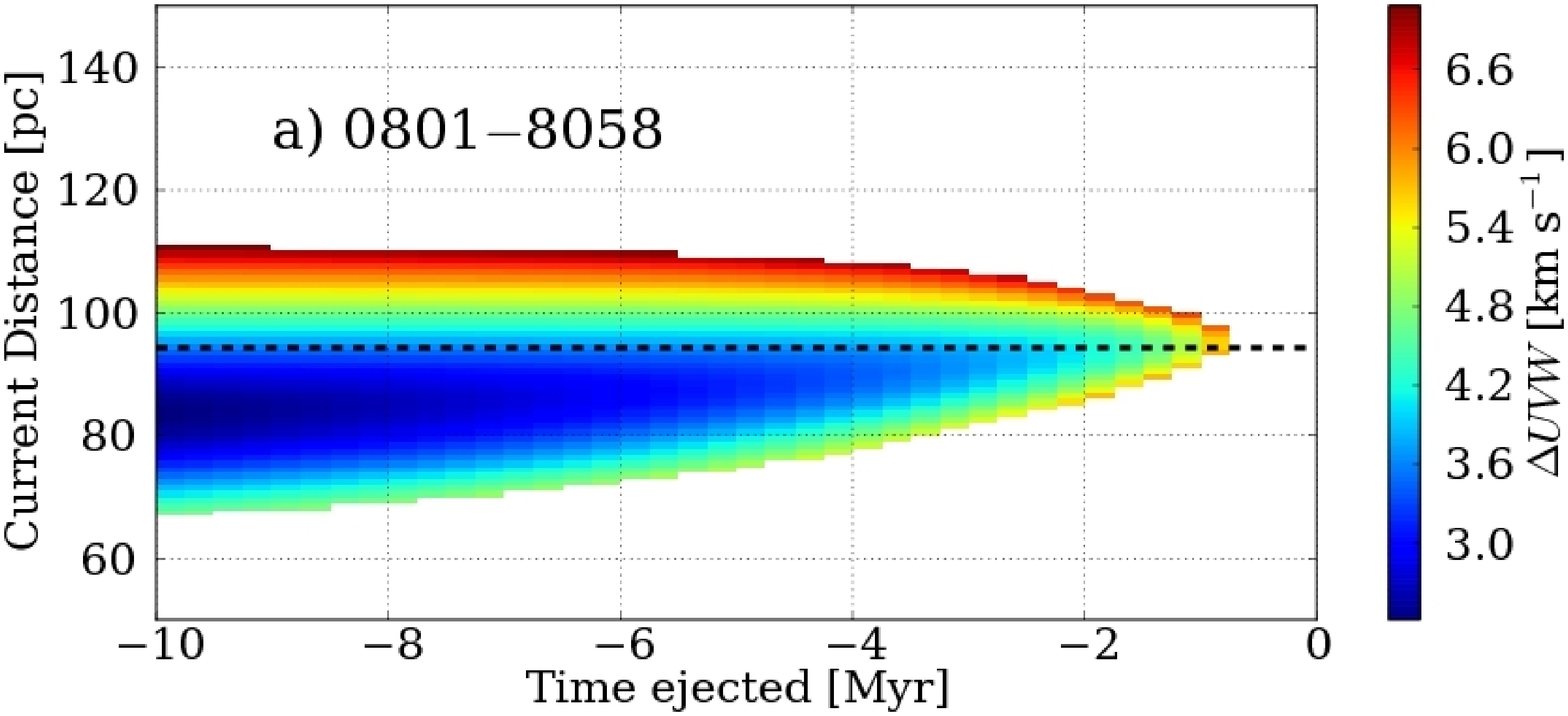} & \includegraphics[width=0.31\textwidth]{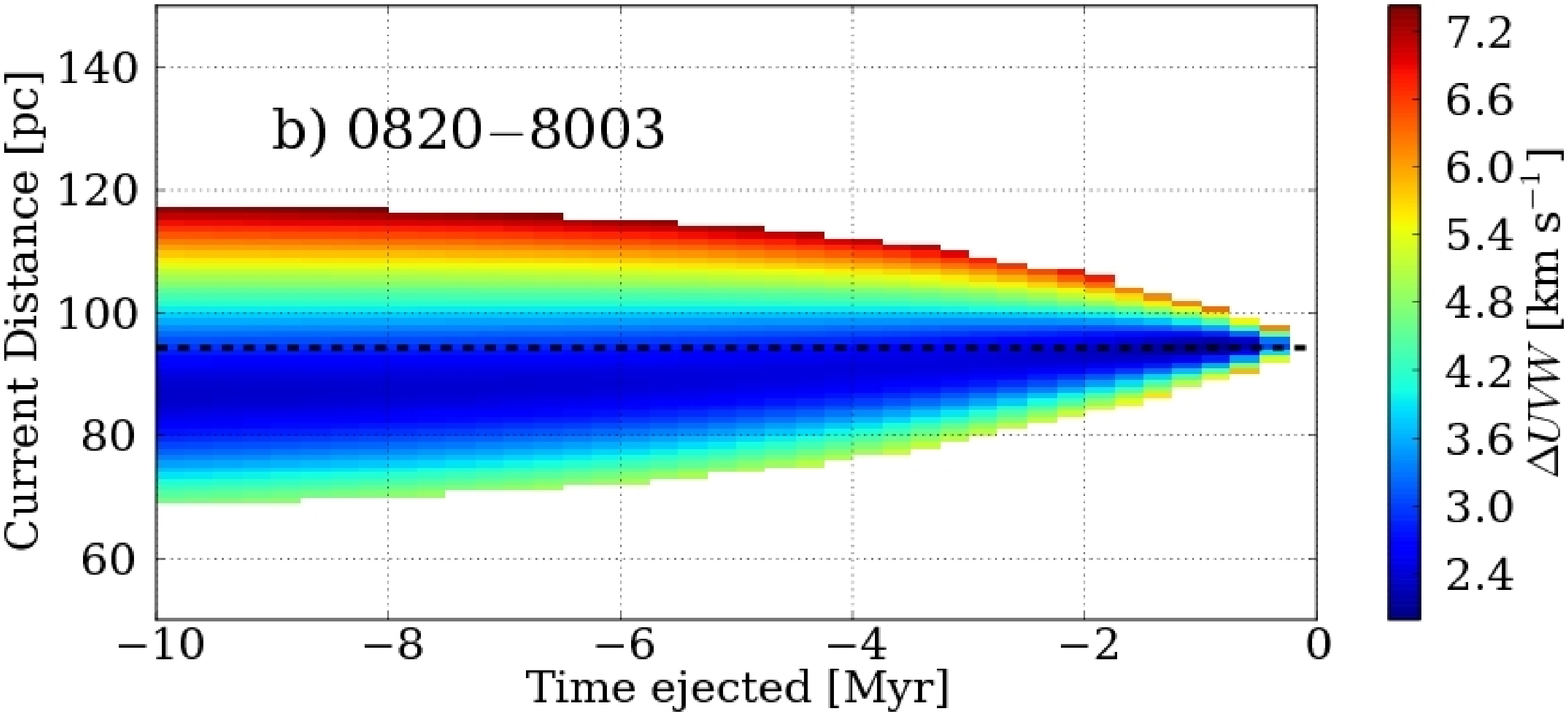} & \includegraphics[width=0.31\textwidth]{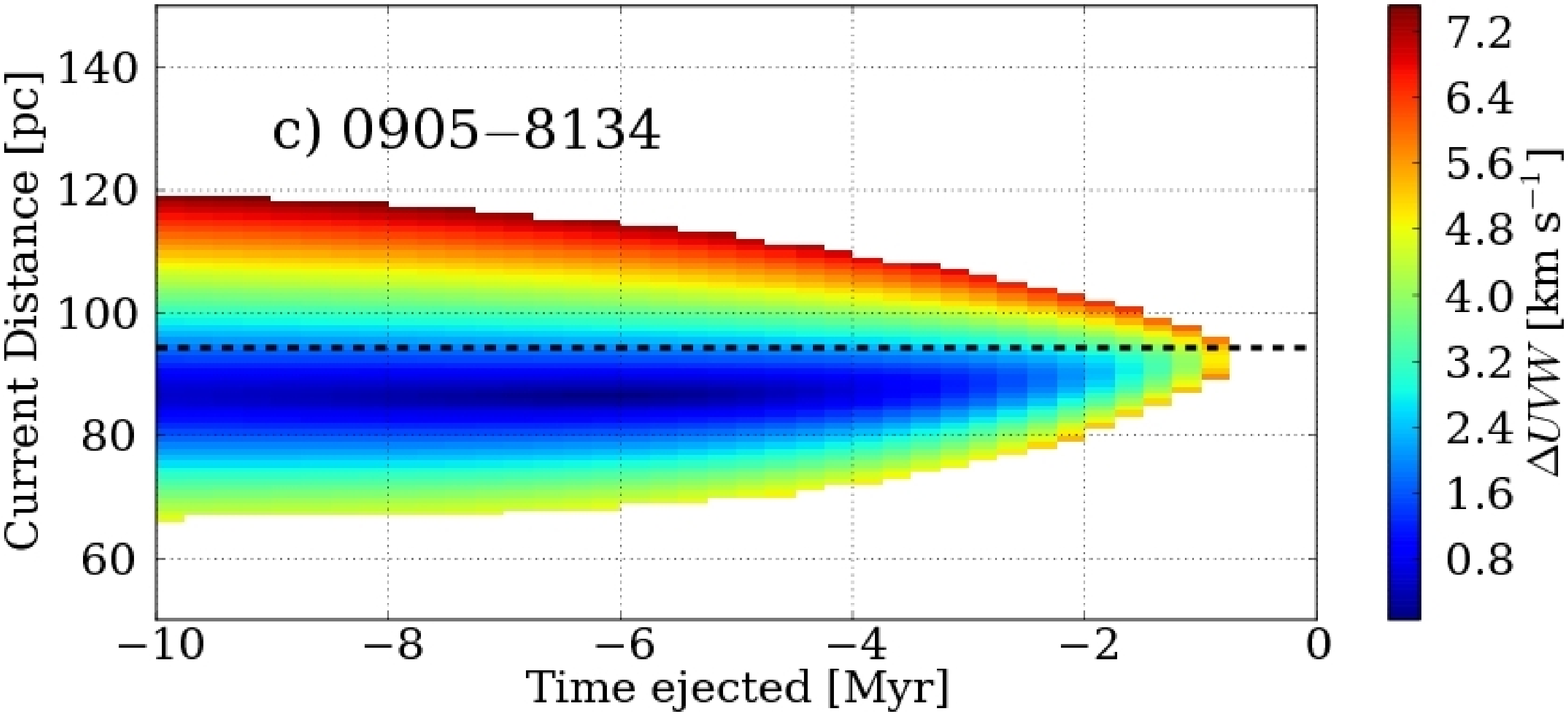}\\ 
 \includegraphics[width=0.31\textwidth]{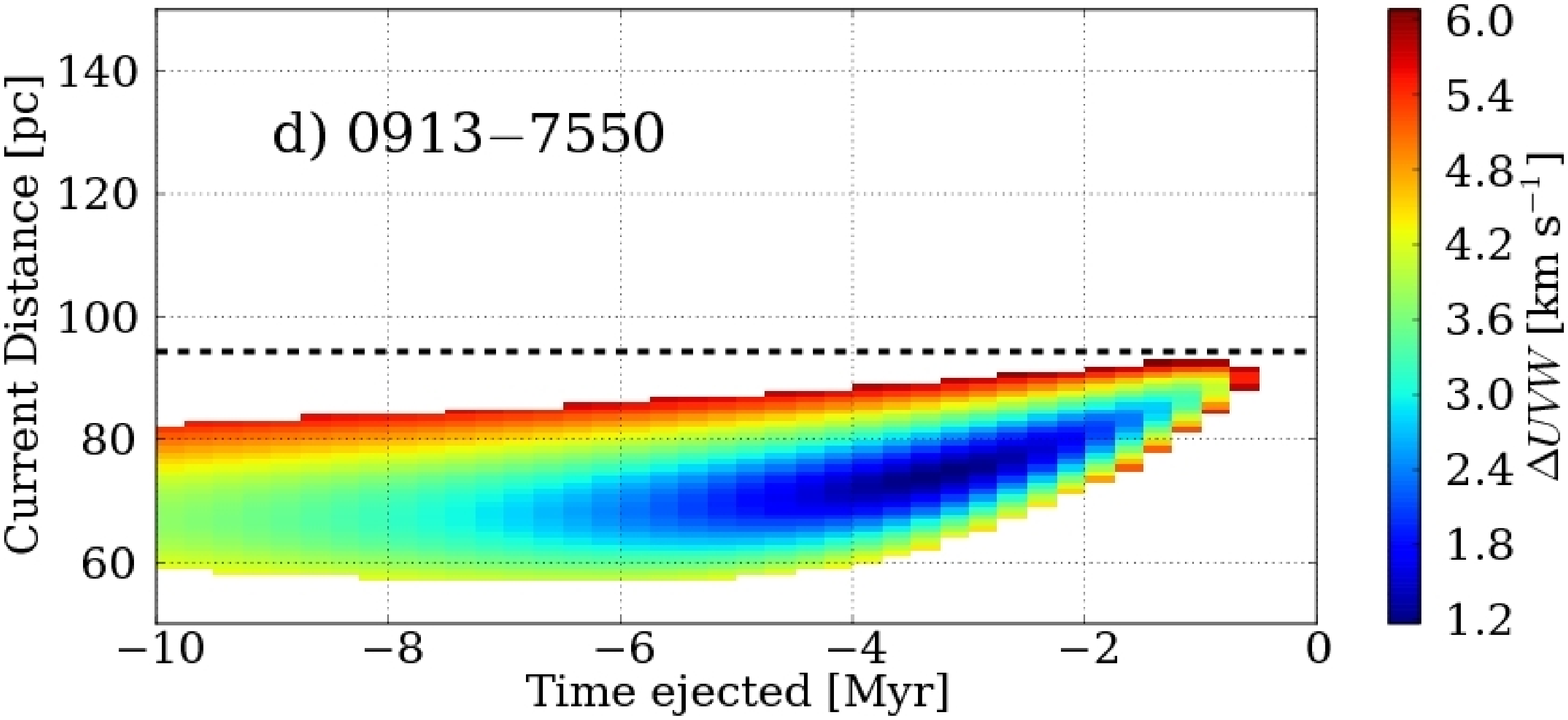} & \includegraphics[width=0.31\textwidth]{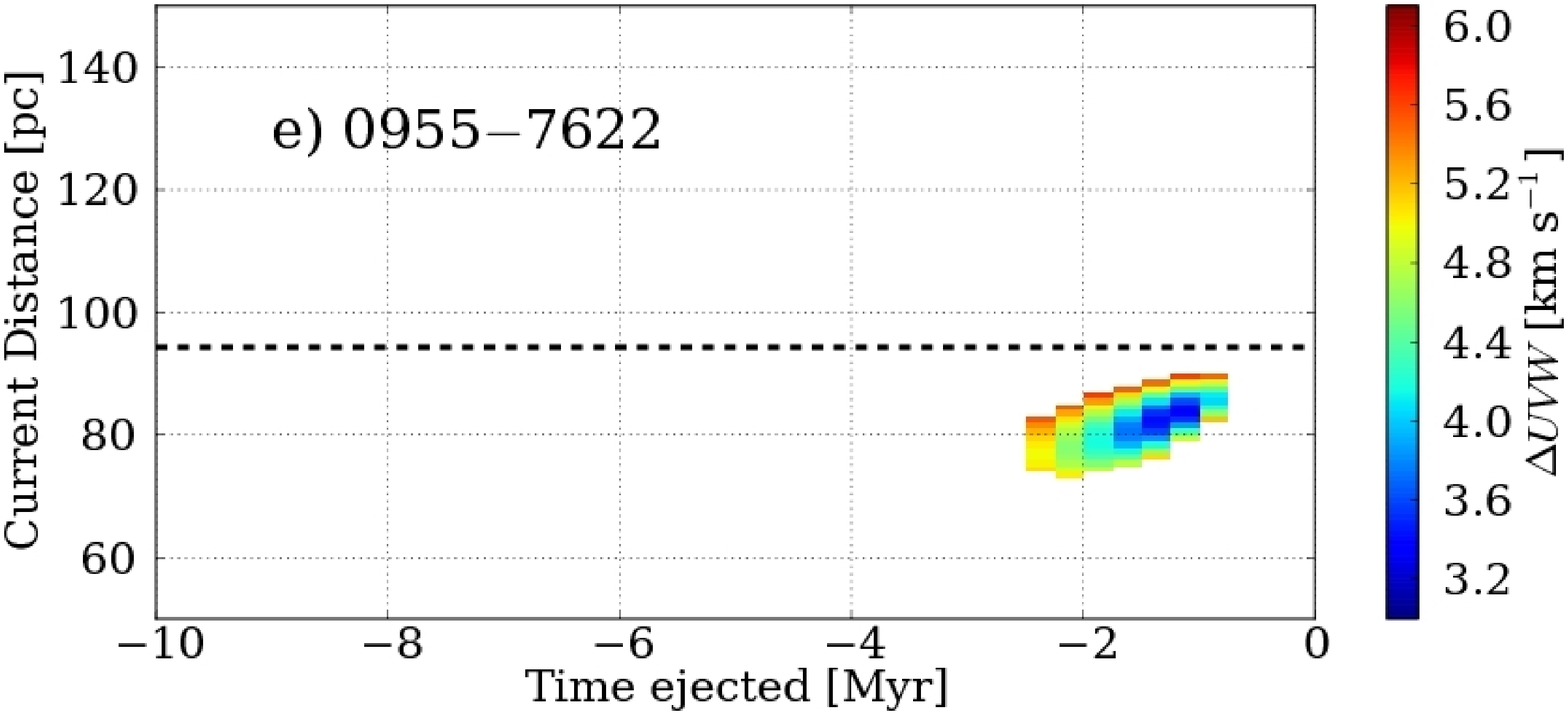} & \includegraphics[width=0.31\textwidth]{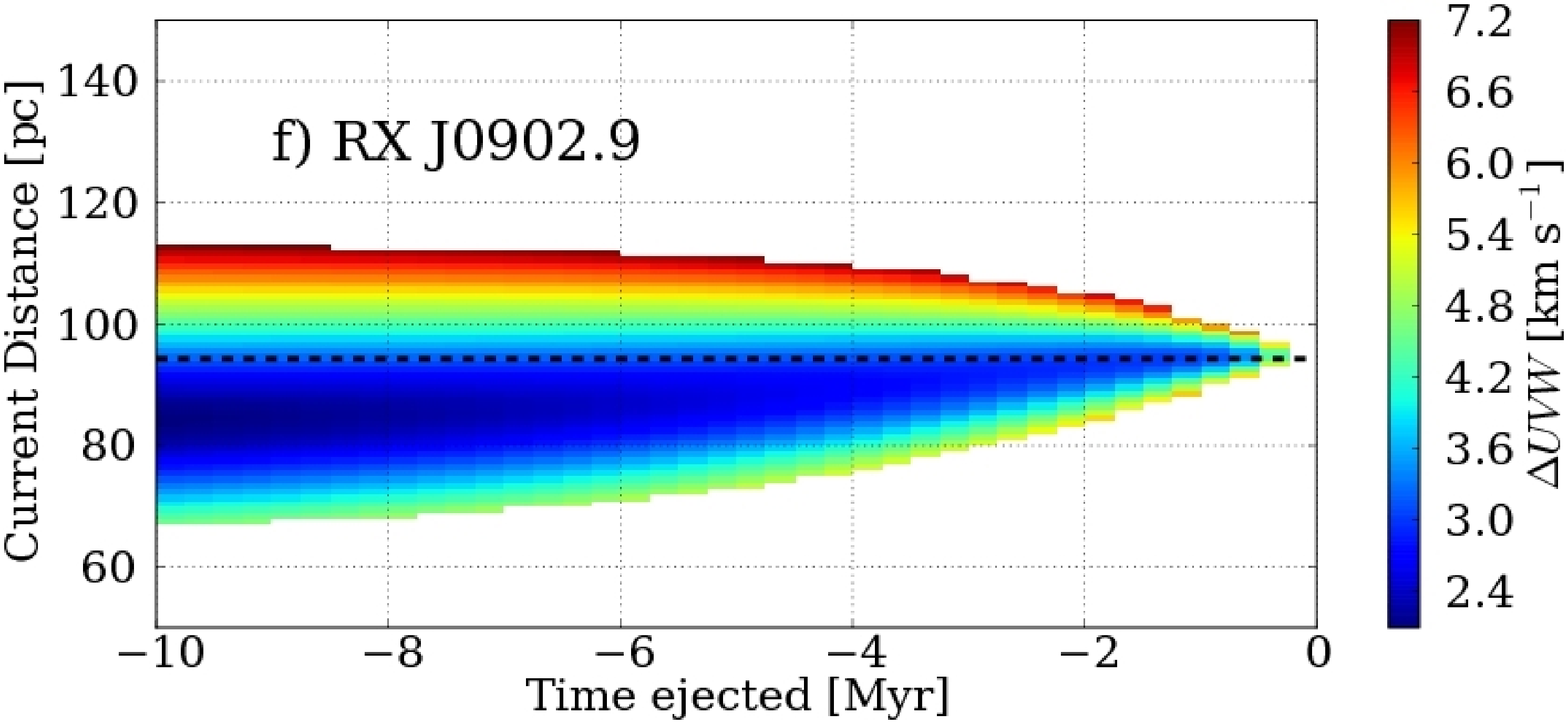}\\ 
 \includegraphics[width=0.31\textwidth]{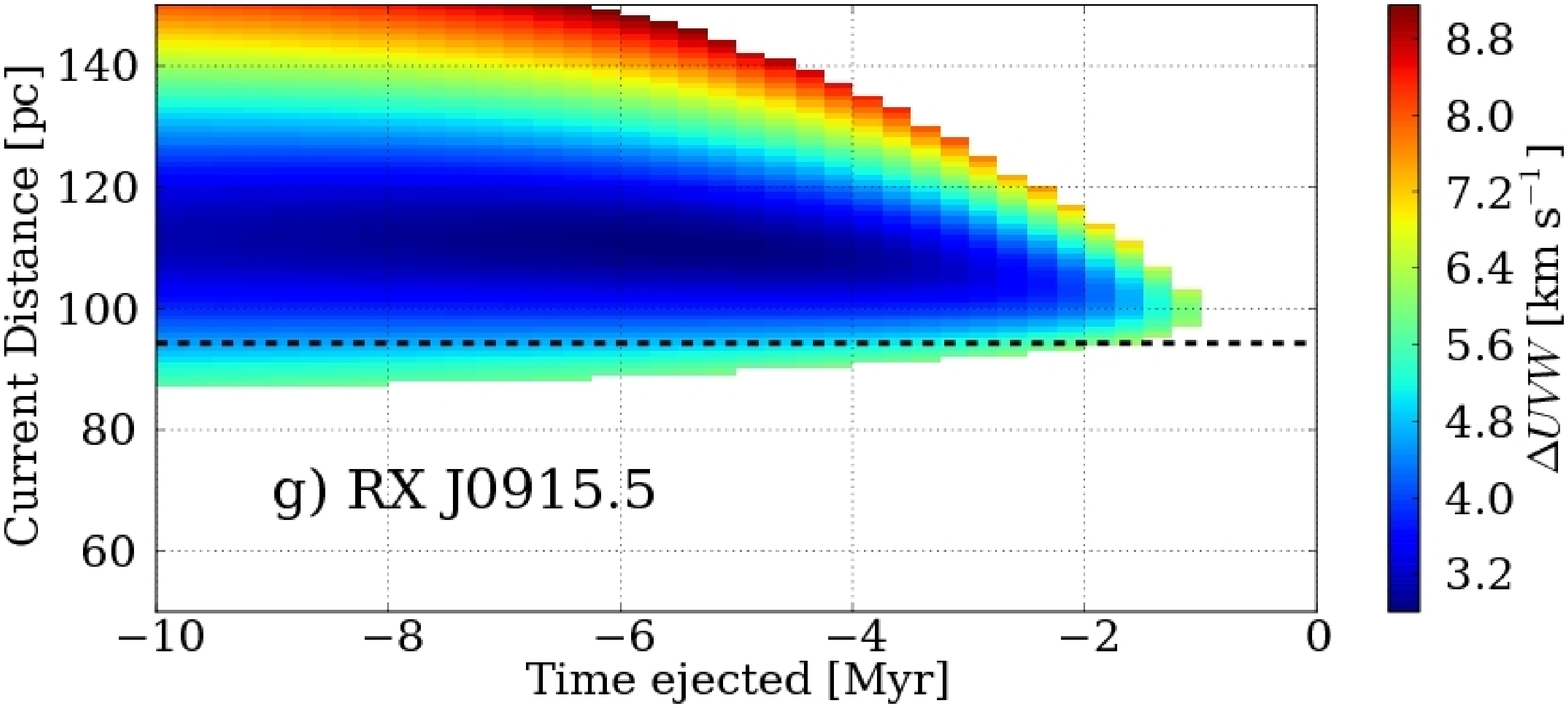} & \includegraphics[width=0.31\textwidth]{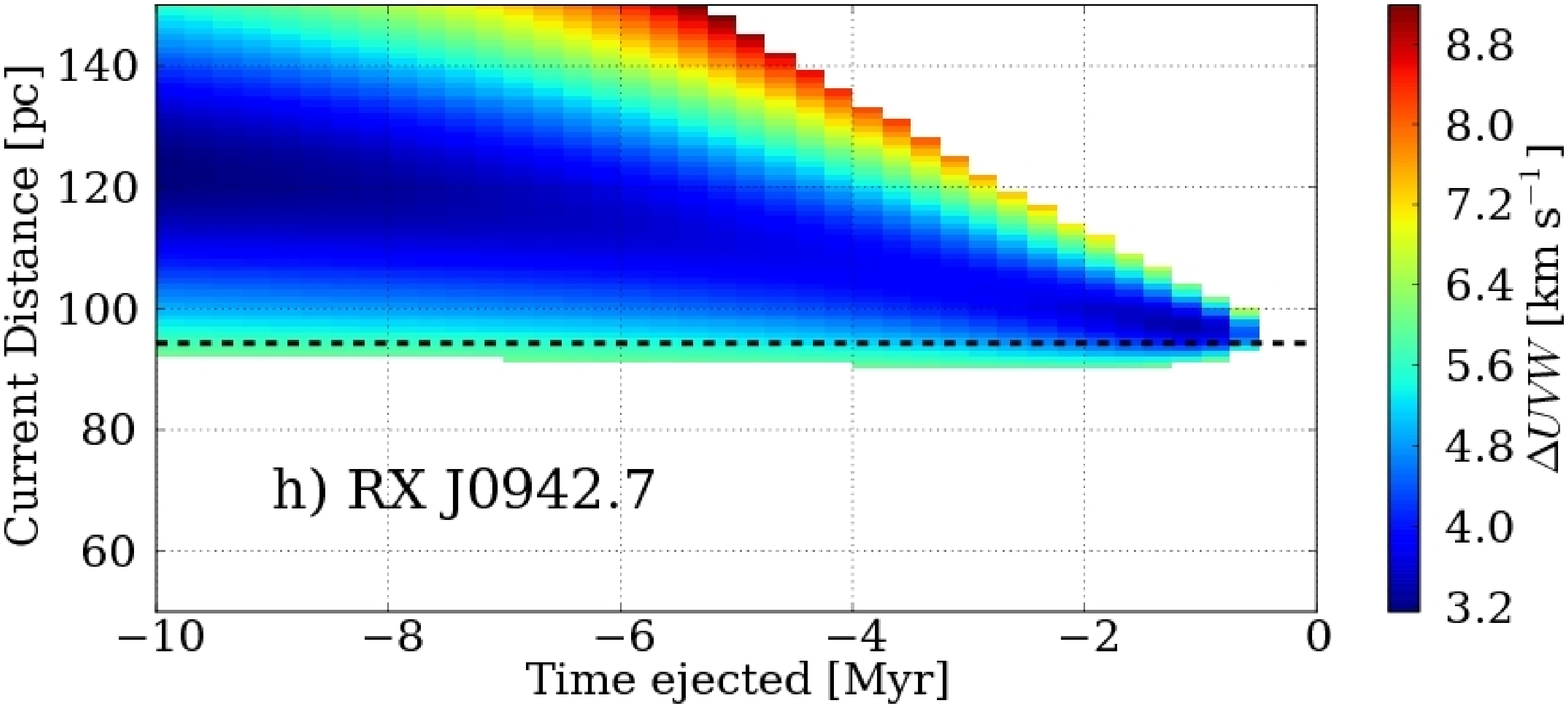} & \includegraphics[width=0.31\textwidth]{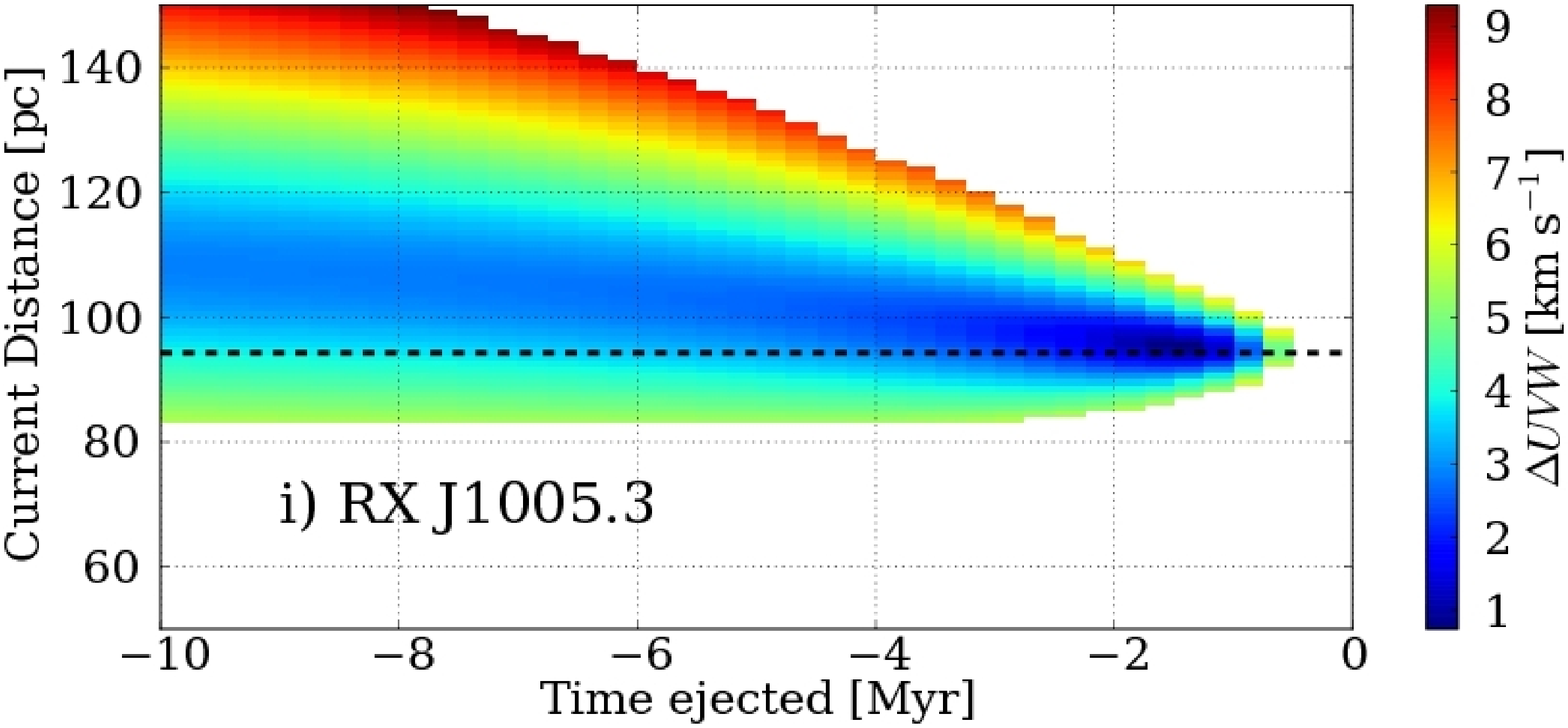}\\ 
   \end{tabular}
   \caption{The results of dynamical simulations for the 5 Li-rich members and 4 PMS stars from \citet{Covino97}. The colour scale indicates the difference between the observed and expected space motion at that distance. The dashed line denotes the revised 94.3 pc cluster distance from \citet{van-Leeuwen07}.}
   \label{fig:deltav}
\end{figure*}

If the remaining candidates are in fact ejected members of \echa\ we do not expect their space motions to be identical to that of the cluster proper. They will have each been imparted some ejection velocity which acts over time to disperse the star away from the cluster core. To determine the possible epoch and magnitude of this impulse we modeled the space motion of the candidates as a function of ejection time and current distance, between $-$10 and 0~Myr, and $50<d<150$~pc. Figure~\ref{fig:deltav} shows the results of such simulations. At each ejection time the cluster was first linearly backtraced to its heliocentric position at that epoch. For each distance we then calculate the current Galactic location of the candidate and the corresponding $UVW$ required to move it there since ejection. This is compared to the observed $UVW$ derived from proper motion and radial velocity measurements. The colour-map in Figure~\ref{fig:deltav} indicates the vector magnitude of the difference between the two space motions. Empty regions in the diagrams have velocity differences larger than those expected from the error in space motion at that distance. A similar map can be made for ejection speed. In these simulations we use the new cluster distance of 94.3~pc and proper motion vector from the revised Hipparcos astrometry of \cite{van-Leeuwen07} and the improved radial velocity, $V_{R}=18.3\pm0.1$~\kms\ (A. Brandeker, private communication). Proper motions from the PPMXL catalogue \citep{Roeser10} are also now available. We use these here and in Table~\ref{table:cand} in preference to the original NOMAD values. Although they quote larger errors than NOMAD, PPMXL is better able to replicate the mean proper motion of known \echa\ members. As a check we have repeated the proper motion anaysis of \S\ref{sec:pm} using the PPMXL values and find no change in the final candidate list. 

\begin{figure} 
   \centering
   \includegraphics[width=0.43\textwidth]{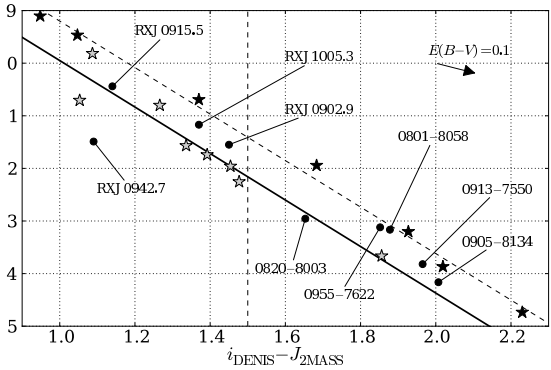} 
   \caption{The late-type $\eta$ Cha cluster sequence, for single (gray stars, solid fitted line) and known binaries (filled stars, dashed line). The arrow indicates a reddening of $E(B-V)=0.1$ mag in this colour-magnitude plane. The vertical line is the blue limit of our photometric/proper motion survey.}
      \label{fig:isochrone}
\end{figure}

The kinematic distances estimated from Figure~\ref{fig:deltav} can be checked against those expected from a star's position in the cluster CMD. Such a diagram is shown in Figure~\ref{fig:isochrone}, where we separate the cluster members into single (or high mass-ratio binary) stars and binary systems. Equal mass binaries should lie 0.75~mag above the single star locus, corresponding to the dashed line. There is excellent overall agreement between the kinematic and photometric distances of the candidates. From comparison of Figures~\ref{fig:deltav} and \ref{fig:isochrone} we propose 4 probable and 3 possible new \echa\ members.

\subsection{Probable members}

\thirtytwo. This candidate is a highly probable member based on its observed space motion. It also has the closest angular separation from the cluster core of any candidate (1.44~deg). The results of the simulations show that it need only be ejected from the cluster at 1--2~km s$^{-1}$ in the last 4~Myr ago to move to its present location. Multiple \textit{WiFeS} observations show the star possesses strong and highly variable H$\alpha$ emission.

\fiftyone. this star shows excellent agreement between its kinematic distance of $\sim$86~pc and slightly elevated position above the \echa\ CMD. The simulations suggest the star was ejected from the cluster 6.25~Myr ago at a speed of 1.5~\kms. 

\fortyone. also shows excellent agreement between kinematic and photometric distances. The star was ejected 3.5~Myr ago at a speed of 6~\kms\ and currently lies in front of the cluster at a distance of 73~pc. Stellar encounters would be so rare and weak at this epoch that the only possible mechanism for such dynamic is from the break-up of an unstable multiple system, perhaps one of the currently seen binaries in the cluster.

RX J0902.9. This M3 star from \citet{Covino97} lies just outside the 1.5~deg surveyed by \citet{Luhman04}. It has a kinematic distance of $\sim$85~pc and would have been ejected at 1--2~\kms\ early in the cluster's dynamical evolution.

\subsection{Possible members}

\twentynine. This star is a probable close binary, based on its location in Figure~\ref{fig:isochrone} and 8 velocity measurements during 2010 January--April. The velocity given in Table~\ref{table:cand} is the mean of these measurements, and should be close to the systemic velocity. The dynamical simulations suggest it was ejected from the cluster at a speed of $<$2~\kms\ early in the cluster's evolution. Even taking into account the apparent binarity, its position in Figure~\ref{fig:isochrone} is more consistent with a distance at or slightly behind the cluster, rather than the 85~pc given by Figure~\ref{fig:isochrone}.

\sixtyfour. Also a possible binary, from its elevated position in Figure~\ref{fig:isochrone} and low velocity. However, ten RV measurements between 2010 January--April do not show any significant variation outside instrument errors. If the simulations are correct, it would have been ejected from the cluster at over 10~\kms\ very late in the cluster's history. Even so, its current distance would still not be close enough to match its position in Figure~\ref{fig:isochrone}.

RX J1005.3. Figure~\ref{fig:isochrone} suggests a distance slightly in front of the cluster, but the best fitting kinematic distance is 95~pc, corresponding to an ejection of 4--5~\kms\ 1.5~Myr ago. It lies at the edge of the 100~$\mu$m IRAS flux in Figure~\ref{fig:skyplot} so it is possible that the star may be reddened slightly (see the arrow in Figure~\ref{fig:isochrone}). 

\subsection {Non-members}

RX J0915.5. The dynamical simulation for this star suggests a distance of 110~pc, whereas its position in Figure~\ref{fig:isochrone} indicates it should lie at a similar distance to the cluster, some 15~pc closer.

RX J0942.7. Not a likely \echa\ member due to its position a magnitude below the cluster isochrone and lower lithium equivalent width. Even at the 120~pc given by the simulations it still lies 0.5 mag below the single-star isochrone. It may be a member of the older, more distant Lower-Centaurus-Crux OB association.

\section{Conclusions}

We have discovered several new probable members of \echa\ up to 5~deg from the cluster core and several possible members awaiting further confirmation. Are these findings consistent with a dynamical origin for the paucity of low-mass objects observed in the cluster? 

To answer this we consider the radial distribution of ejectees presented in \citet{Moraux07}. Integrating the distribution out to 9~pc (5.5 deg) we could expect to find up to 6 stars across all masses in the survey area. Our blue photometry limit of $i-J>1.5$ corresponds to an approximate spectral type of M3, whereas our spectroscopic campaign is complete to  $i-J<2$, approximately M5. Transforming these spectral types to temperatures using the relation of \cite{Bessell91} and then to masses from an 8~Myr \citet{Baraffe98} isochrone gives an \emph{approximate} surveyed mass range of $0.08<M<0.3$~\msun. \citeauthor{Moraux07} find the mass distribution of ejectees is roughly constant with radius and consistent with the \citet{Chabrier03} input IMF. We therefore integrate this IMF over the above mass range to find the fraction of stars at those masses, this is $\sim$40\%. Hence we can expect to find 2--3 bona fide \echa\ members within the surveyed area and mass range. Our discovery of 3 probable 2MASS/DENIS members is consistent with this prediction. We can therefore conclude that dynamical evolution is solely responsible for the current configuration of \echa\ and it is not necessary to invoke an IMF deficient in low--mass objects.

\end{document}